# Local Whole-Device Scanning of Distortion in Superconducting Microwave Resonators

S.K. Remillard, *Member, IEEE*, A.E. Wormmeester, and R.A. Huizen

*Abstract*— Using a near-field microwave technique, two-dimensional images have been made of the second and third order intermodulation distortion (IMD) of $Tl_2Ba_2CaCu_2O_8$ and $YBa_2Cu_3O_7$ thin film microwave resonators. It was found that second and third order IMD do not have identical spatial distributions, which indicates that physical mechanisms play different roles in their generation. This technique enables the investigation of the roles of these mechanisms. As an illustration, the sensitivity to magnetic fluxons of the two orders is described, with second order being more sensitive to fluxon density.

*Index Terms*— Intermodulation distortion, Superconducting microwave devices, High temperature superconductors

## I. INTRODUCTION

Microwave properties of superconductors are usually found globally using quality factor (Q), resonant frequency, and distortion. Resonant frequency is related to the kinetic inductance $L_K$, and Q is related to the surface resistance $R_S$, both averaged over the superconductor. Distortion generated throughout the superconductor is usually measured as an average over the entire sample. But distortion is presumably generated in locations where the microwave current is either high or influenced by a defect. The purpose of this paper is to show that intermodulation distortion (IMD) can be measured locally, resulting in the first whole-device scans of microwave nonlinearity at multiple orders. An investigation into the role of fluxons on even and odd order nonlinearity will illustrate the effectiveness of this technique in investigating fundamental physical mechanisms of nonlinearity.

Prior works on local nonlinearity have employed a scanning probe to detect electric field radiated at the IMD frequency [1], near-field microwave microscopy to examine harmonic distortion [2], and laser scanning microscopy to examine third order IMD [3]. This paper adds to these tools by describing the simultaneous local measurement of second and third order nonlinearity at the same frequency.

Nonlinearity is seen in the surface impedance $Z_S(K)=R_S(K)-j\omega L_K(K)$, where $\omega$ is angular frequency, and $K$ is surface current density which drives the nonlinear behavior [4]. However, distortion is detected at lower microwave current than that which affects $Z_S$. At the high temperatures of this work, distortion usually increases with the expected slope (3:1 for third order [5], 2:1 for second order) for several decades of input power before $Z_S$ starts to change. The current dependence of $Z_S$ is not an agent of nonlinear emissions. As with distortion, it is a consequence of several mechanisms.

The order of the distortion corresponds to the generating term in the Taylor series expansion [6]. Different mechanisms can dominate even and odd order distortion. In the resistively and capacitively shunted junction model, current dependent conductance of grain boundaries can contribute to third order nonlinearity [7]. Sometimes only seen in a static magnetic field [8], second order nonlinearity is evidence of time reversal symmetry breaking [9], which could have origins in the superconducting state physics, such as Andreev reflections at superconductor boundaries [10]. Above 77K, where the results in this paper were collected, fluxon intrusion at the film edge contributes to both orders [11]. The technique described below shows that second order nonlinearity at elevated temperature more closely follows the density of fluxons than does third order [12].

Both orders reach a peak just below the critical temperature, $T_C$ [8]. Second order nonlinearity is usually less peaked in temperature near $T_C$ than is third order. In Ref. [13] it was reported that the third order IMD of $YBa_2Cu_3O_7$ was undetectable below 82 K, but the second order IMD was strong down to, and presumably well below, 77K. In order to enhance the IMD, the data presented here were taken near $T_C$.

Intermodulation distortion, which in superconductors has the same physical root cause as harmonic distortion of the same order [14], is a concern in passive microwave devices due to the spurious tones generated inside of the operating band. In this work we mapped the local second and third order IMD generated inside superconducting resonators which were designed for use in 800 MHz wireless receive filters.

## II. WHOLE DEVICE SCANNING OF LOCAL NONLINEARITY

### A. Intermodulation Distortion

Details of the measurement are provided in Refs. [15,16]. Briefly, the sample is driven at its resonant frequency $f_3$. Two low frequency tones $f_1$ and $f_2$ are also introduced to the

Manuscript received September 9, 2017; Revised manuscript receieved December 23, 2017. This material is based in part upon work supported by the National Science Foundation under Grant Numbers DMR- 1206149 and DMR- 1505617. Any opinions, findings, and conclusions or recommendations expressed in this material are those of the authors and do not necessarily reflect the views of the National Science Foundation.
 (Corresponding author: S.K. Remillard.)
 S.K. Remillard, A.E. Wormmeester, and R.A. Huizen are with Hope College, Holland, MI, 49423 USA. (e-mail: remillard@hope.edu).

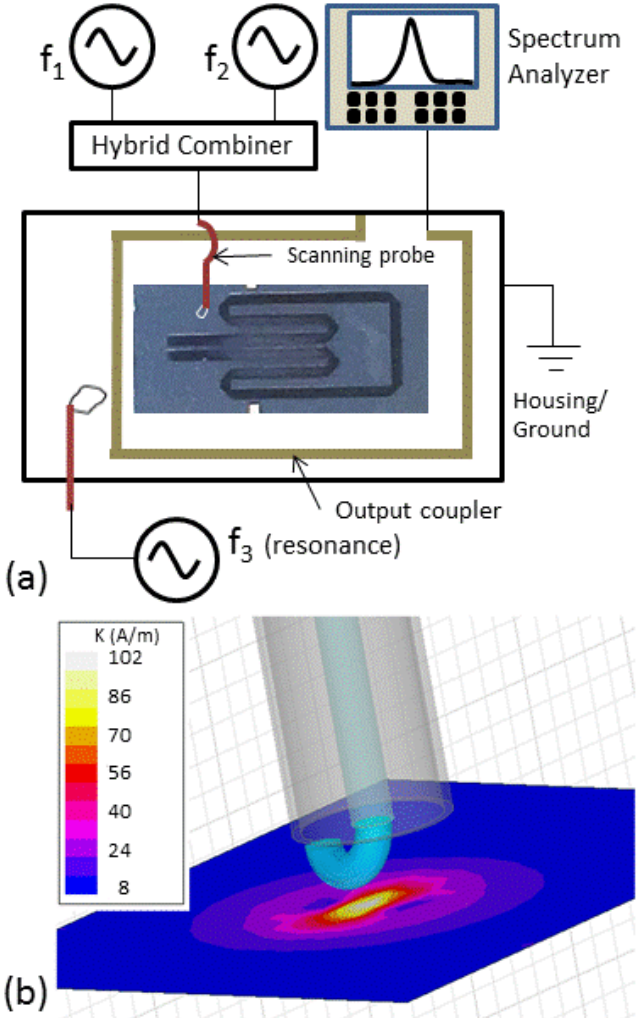

Fig. 1. (a) Configuration for IMD. Two "probe" signals at $f_1$ and $f_2$ are combined externally and sent into the resonator, which is excited by $f_3$. The resulting IMD is picked up by the spectrum analyzer. (b) A simulation using *Ansys® HFSS Release 13.0* shows the locally induced surface current $K$ from the probe.

resonator. These frequencies are smaller than the half-width $\delta f$ of the resonance. For example, if the loaded $Q$ $(=f_3/\delta f)$ for a 1 GHz resonator is 1,000, then $f_1$ and $f_2$ both need to be less than 500 KHz. Second order IMD occurs in the passband at $f_3 \pm f_1$ and $f_3 \pm f_2$. Third order IMD is at $f_3 \pm |f_2 - f_1|$. Even and odd order IMD can then be measured at nearly the same frequency.

Being out of band, $f_1$ and $f_2$ do not propagate through the device under test (DUT). Introducing $f_1$ and $f_2$ with a probe in proximity of the sample induces near-field current as shown in Fig. 1b. Mixing between the near-field current and the current of the resonant mode at $f_3$, which is excited by a separate coupler, produces IMD locally at the point of the probe. The key concept of this test is for the distortion to reside within the passband of the DUT, which then excites its resonance in order to be detected. In Ref. [17], the probe moved along a section of a hair-pin resonator producing a 1-dimensional map of the IMD that corresponded primarily to current in the structure. An IMD hot-spot at an engineered defect was clearly visible. This method was also benchmarked against photoresponse IMD.

Because mixing only occurs where the probe induces local current, the three-tone IMD method is robust and free of "set-up" IMD. The same cannot be said here for harmonic distortion, where the signal source can produce harmonics as large as -50 dBc, requiring a filter for the signal generator with at least 100 dB of rejection at the resonant frequency of the DUT, as was done in Ref. [18], but with a non-resonant DUT. Signal generator harmonics would pass through the DUT at the exact frequency as the generated harmonics. Were it not for this complication, a simpler test would involve a single tone at $f_3/2$ or $f_3/3$ introduced by the movable probe.

*B. The Probe*

The two low frequency probe signals are magnetically coupled to the superconducting surfaces with a small loop fashioned from semi-rigid coaxial cable. The probe for two of the scans was made from 0.047 semi-rigid copper cable allowing for a loop, formed from the center conductor, with an inner diameter of about 320 μm. In order to further localize the field, a copper shield surrounded the loop. Efforts are currently underway to use 0.020 semi-rigid cable to form a 150 μm loop, as well as using a magnetic circuit probe as was employed in the near-field microscopy of Ref. [19].

The probe is mounted onto a Lakeshore Cryotronics model MMS-07 Micro Manipulated Stage, with 3-dimensional micrometer controlled motion. The sample is cooled by a CTI model M22 coldhead, which also cryopumps the roughing pump backed vacuum chamber. Signals $f_1$ and $f_2$ enter through the probe which is located by the coordinates $(x, y)$ in the scans that follow. The probe height above the sample is approximately 20 μm. This probe height is set for each scan to the lowest height that preserves the stability of the $f_3$ signal.

III. WHOLE DEVICE SCANNING RESULTS

*A. Straight line resonator*

Each resonator had been made earlier by argon ion beam milling. A 150 μm wide, 27 mm long, straight transmission line was fabricated from a 400 nm thick $Tl_2Ba_2CaCu_2O_8$ film on a 0.5 mm thick $LaAlO_3$ substrate coated on both sides. A gold film was deposited on top of the ground plane to facilitate indium solder attachment to a gold plated titanium plate.

For this scan only, the probe was made from 0.086 semi-rigid cable, which has a 500 μm diameter center conductor, allowing for a loop with a 590 μm inner diameter. The probe was driven at 1 mW (0 dBm). The resonator was excited in its second harmonic mode at 2.4 GHz at temperature 100 K, where $T_C$=105 K. Unloaded Q at this temperature was about 500. Two current maxima thus occur between the midpoint of the line and its ends. The second and third order IMD are shown in Fig. 2, and for both orders the IMD is large where the current is large, although there is more third order nonlinearity at the anti-node located at y=22 mm.

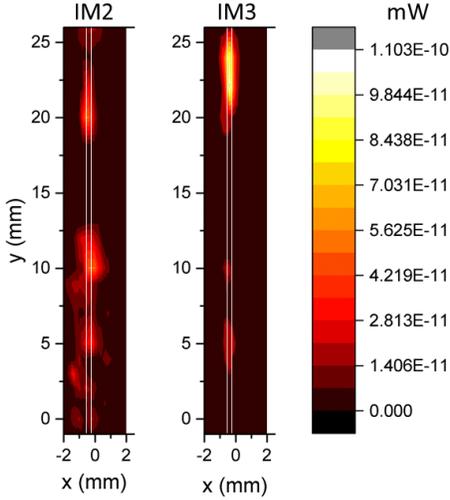

**Fig. 2.** Second (IM2) and third (IM3) order IMD generated in a straight-line resonator excited in its second harmonic mode driven at 0 dBm. The white outline shows the 27 mm long resonator's location, which starts around y=1mm. Plotted using Origin (OriginLab, Northampton, MA).

*B. Folded hairpin resonator*

Similarly fabricated as the previous sample, but with the intention of reducing nonlinearity, the $Tl_2Ba_2CaCu_2O_8$ folded hairpin resonator in Fig. 3 had 1.0 mm linewidths. The scan, using the probe described in Section IIB above, was performed at 91 K and the $T_C$ was 101 K. The probe was raster scanned with step sizes of 50 μm to 200 μm.

Operated in its fundamental harmonic mode, the current is highest along the bottom portion at y= -5.5 mm. Simulated current distribution (not shown) reveals the expected points of current crowding around inner corners which in fact lead to another high current region at the top end at y=5 mm.

Two-dimensional raster scans of the second and third order IMD are shown in Fig. 3. There is a hot spot in the second order IMD in the top left of the scan at y=5.5 mm, which has been reproducible after each step in the evolution of the probe design. Some IMD can be expected there due to current crowding at the bend. However, the second order in this location is much higher than it is elsewhere, even though this is not the point of highest current.

*C. Narrow line folded hairpin*

A single resonator made from 400 nm thick $YBa_2Cu_3O_7$ on a 0.5 mm thick $LaAlO_3$ substrate and with a fundamental resonant frequency of 835 MHz was scanned at 86 K and an input of +10 dBm. The $T_C$ was 91 K. The transmission line width was 200 μm with line spacing of 150 μm.

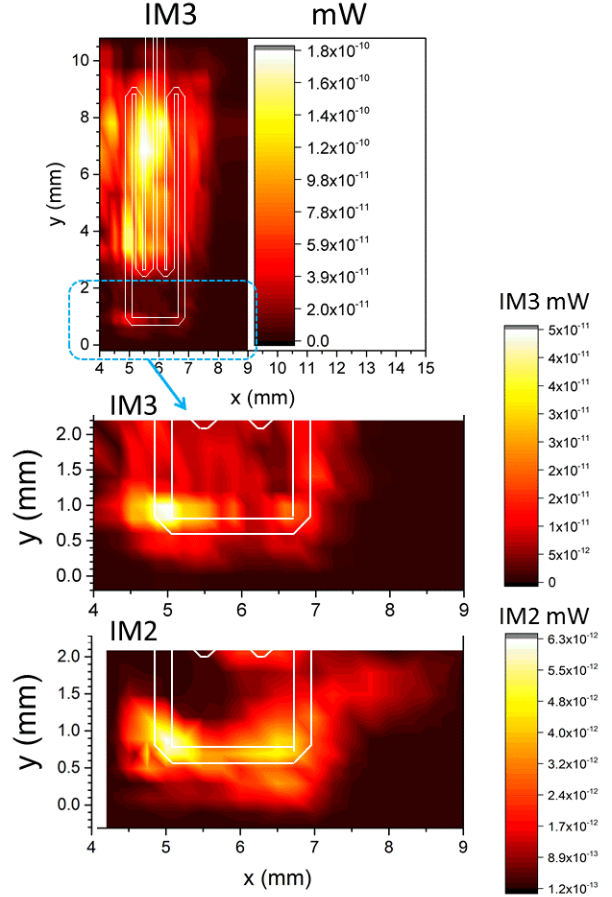

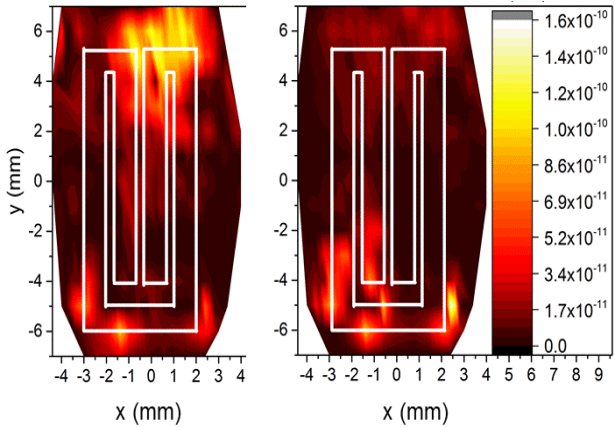

**Fig. 3.** Second order (left) and third order (right) IMD in a folded hairpin resonator with 1 mm wide lines. The outline of the resonator is superimposed. There is a hot spot in the second order IMD in the upper right corner.

**Fig. 4.** IMD scan of the 835 MHz narrow-line folded hairpin resonator at 86K excited in its fundamental mode. The close-up view shows that the left corner emits more third order IMD than the right corner.

Above y=2.5 mm the lines are very closely separated, by approximately 100 μm, and the IMD signal is very high. Although the current density becomes weaker away from the center point of the folded microstrip line, the loop probe, which is larger than the line spacing, couples to multiple lines at y>2.5 mm. A close-up view of the region below y=2.0 mm reveals the IMD being produced strongly from the corners of the folded-structure, and more strongly from the left corner at x=5 mm, especially for the third order.

## IV. DISCUSSION AND APPLICATIONS

The movable probe induces nearfield current, which then generates nonlinearity by mixing. The nonlinearity is in-band of the DUT and thus becomes a source of excitation, which is then picked up by a fixed coupler which most conveniently encircles the entire device. Opportunities are opened up by this capability, which are illustrated by these and other recent results. Those explored thus far include searching for nonlinearity hot spots [17], understanding the distinction between second and third order nonlinearity [ 20 ], and modelling the contribution of fluxons to nonlinearity [12].

Differences can be seen in Figs 2, 3, and 4 between the distributions of second and third order IMD. Close to $T_C$, both orders correspond to field modulation of the penetration depth through the nonlinear Meissner effect, but second order nonlinearity also requires time reversal symmetry breaking, which then connects second order nonlinearity to dissipation. At high current, both orders become saturated, *i.e.* current independent. The two orders saturate at neither the same current nor at the same temperature [20]. Second order saturates due to complete decoupling of weak link grain boundaries. Third order nonlinearity saturates because the penetration depth has become larger than the skin depth [20]. For the scan in Fig. 4, the third order was saturated, which might explain the hot spot at x=5 mm for third order. Nonlinearity hot spots emerge from material defects [21] and they do so differently for the two orders, as seen in Fig. 3. Ref. [17] investigated hot spots in third order IMD due to a defect engineered by heavy ion beam irradiation. Both Refs. [17] and [21] used photoresponse to generate highly resolved third order IMD images. The currently reported technique enables synchronous comparison of the two orders.

As in Ref. [12], in order to examine the roles of fluxons, a 230 Gauss static magnetic field was briefly applied perpendicularly to the narrow-line folded-hairpin resonator near $T_c$. Upon removal of the field, second and third order IMD relax, but not in the same manner. In Fig. 5, the third order relaxes logarithmically with time elapsed since removal of the field, as expected from the Bean-Livingston model, as was also observed in the global measurements of Ref. [22]. The second order fits two processes, one fast and exponential and one slow, logarithmic, and more pronounced than third order. The fast process is remanent magnetic relaxation [23] and only occurs in second order. Closer to $T_C$ the fast decay is even faster, perhaps because the fluxon viscosity becomes very small near $T_C$ [24]. As this work now progresses, rastering in a static field is enabling the investigation of the role of extrinsic fluxon-based nonlinearity in the hot spots.

In order to improve resolution, a 100 μm diameter loop probe will be used next. The orientation of the probe loop may also influence results since locally induced current is perpendicular to the dipole moment of the vertical probe. Scans performed with the probe oriented either transverse to or along the transmission line did reveal occasional differences. However, the IMD hot spots were manifest regardless of how the probe was oriented. To negate this potential effect, an omnidirectional loop oriented parallel to the transmission line will be investigated.

The scan in Fig. 3 was made on the center resonator of a three pole filter. The other two resonators were disabled by sputtered gold. With the concept proven for single resonators, whole device scans of multi-section filters are possible, providing more useful design guidance. IMD does not originate equally from each resonator in a filter, and as the first design implication of this technique, it was shown in Ref. [15] that what is perhaps the most problematic IMD in a wireless front-end, that due to out-of-band carriers, is effectively mitigated by eliminating IMD from the input resonator. Additionally, the fluxon time-decay of resonator samples used in magnetic, electron [25], or other quantum spin resonance [26] can now be imaged.

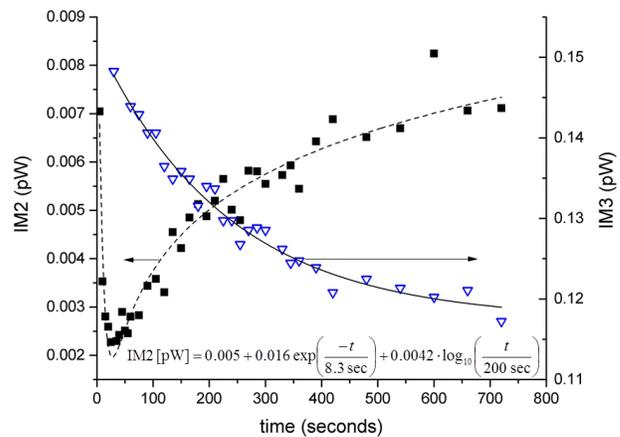

**Fig. 5.** Second (dashed) and third (solid) order IMD versus time since removal of a static magnetic field for the YBa$_2$Cu$_3$O$_7$ narrow-line folded hairpin resonator at 88K revealing, through the fits, the influence of remanent magnetic relaxation on second order IMD, and the Bean-Livingston model on both orders. Fit using Origin (OriginLab, Northampton, MA)

## V. CONCLUSION

Two-dimensional whole device scans of locally generated IMD reveal hot spots in the second and third order nonlinearity. The even and odd order nonlinearities depend to different extents on several root causes associated with kinetic and dissipative effects. The development of whole device IMD imaging reported here is beginning to advance the field through its application to defect analysis, synchronous even and odd order nonlinearity comparison, and investigation into the effect of fluxons on spatially localized nonlinearity.

## ACKNOWLEDGMENT

Earlier contributions to this project were made by Sean Hamilton and Alec Nelson.